# On the channel width-dependence of the thermal conductivity in ultra-narrow graphene nanoribbons

Hossein Karamitaheri[1] and Neophytos Neophytou[2*]

[1]Department of Electrical Engineering, University of Kashan, Kashan 87317-53153, Iran

[2]School of Engineering, University of Warwick, Coventry, CV4 7AL, UK

[*]N.Neophytou@warwick.ac.uk

## Abstract

The thermal conductivity of low-dimensional materials and graphene nanoribbons in particular, is limited by the strength of line-edge-roughness scattering. One way to characterize the roughness strength is the dependency of the thermal conductivity on the channel's width in the form $W^{\beta}$. Although in the case of electronic transport this dependency is very well studied, resulting in $W^6$ for nanowires and quantum wells and $W^4$ for nanoribbons, in the case of phonon transport it is not yet clear what this dependence is. In this work, using lattice dynamics and Non-Equilibrium Green's Function simulations, we examine the width dependence of the thermal conductivity of ultra-narrow graphene nanoribbons under the influence of line edge-roughness. We show that the exponent $\beta$ is in fact not a single well-defined number, but it is different for different parts of the phonon spectrum depending on whether phonon transport is ballistic, diffusive, or localized. The exponent $\beta$ takes values $\beta < 1$ for semi-ballistic phonon transport, values $\beta \gg 1$ for sub-diffusive or localized phonons, and $\beta = 1$ only in the case where the transport is diffusive. The overall $W^{\beta}$ dependence of the thermal conductivity is determined by the width-dependence of the dominant phonon modes (usually the acoustic ones). We show that due to the long phonon mean-free-paths, the width-dependence of thermal conductivity becomes a channel length dependent property, because the channel length determines whether transport is ballistic, diffusive, or localized.

**Keywords:** low-dimensional phonons, graphene nanoribbons, thermal conductance, line-edge roughness, atomistic simulations, non-equilibrium Green's functions - NEGF.

Graphene nanoribbons (GNRs) and low-dimensional materials in general, have recently attracted significant attention, both for fundamental research as well as for technological applications [1-14]. Specifically with regards to their thermal transport properties, these materials are heavily studied as they exhibit features distinctively different from bulk materials such as deviation from Fourier's law [3, 6, 10, 15, 16], crossover from ballistic into diffusive transport regimes [17, 18], even a counter-intuitive increase in the thermal conductivity with confinement [19], and extremely high thermoelectric properties [12].

The major effect in limiting thermal conductivity in 1D channels, however, seems to be boundary scattering [7, 20, 21], and thus, large efforts are devoted into understanding the influence of surface roughness, and line-edge roughness in particular for GNRs, on the thermal conductivity. One of the most common ways to characterize the effect of roughness on the conductivity of the channel and determine the transport regime of operation, is the use of specific power-laws in the form of $L^{\alpha}$ and $W^{\beta}$, where $L$ is the length of the channel, and $W$ is the width of the channel. The length-dependence power-law on the thermal conductivity has been studied extensively [22-25]. With regards to the *width-dependence*, a lot of work can be found for *electronic transport*, yielding under diffusive transport conditions $W^{6}$ for nanowires [26- 28], and $W^{4}$ for nanoribbons [29]. In the case of *phonon transport*, however, no systematic work to-date exists that studies the width-dependence of the thermal conductivity of low-dimensional channels in the presence of line-edge roughness.

In this work we perform such an investigation using simulations of phonon transport in ultra-narrow GNRs by employing lattice dynamics for the phonon spectrum and the non-equilibrium Green's function (NEGF) approach for transport. The method is described in detail in our previous work, where we explored the length-dependence of the thermal conductance of GNRs [25, 30]. We show that the exponent $\beta$ is in fact not a single well-defined number, but it takes different values depending on whether the phonon transport is quasi-ballistic ($\beta < 1$), diffusive ($\beta = 1$) or localized ($\beta \gg 1$).

The 4[th] nearest-neighbor force-constant-method that we use (with parameters from [30]) can correctly regenerate the bandstructure of graphene as compared to experimental data [12], and also provides very good agreement with experimental data for rough GNRs with widths up to $W \sim 15$ nm [31]. Figure 1a shows for reference a typical phonon spectrum for a GNR channel of width $W = 4$ nm with the colormap showing the contribution of each phonon state to the ballistic thermal conductance at room temperature, (light indicates high and dark indicates low contribution). Using NEGF we then simulate phonon transport in GNRs of widths $W = 1$ nm up to $W = 5$ nm and lengths from $L = 5$ nm up to $L = 1000$ nm. To construct the line-edge-roughness geometry we use an exponential autocorrelation function with root mean square of the roughness amplitude $\Delta W_{rms} = 0.1$ nm and roughness correlation length $\Delta L_C = 2$ nm [31]. We extract the frequency dependent phonon transmission, $T_{ph}$, and normalize it by the channel width $W$. The quantity $T_{ph}/W$ will be referred to as the 'normalized transmission', and has the same width-dependence as the thermal conductivity. For all data we average over 50 different roughness realizations.

The transmission function in the ballistic regime is equal to the number of modes at given energy/frequency. This Landauer approach is widely used for phonon and electron transport in nanostructures. In the presence of scattering, the contribution of each mode to transport is reduced, which is captured by the reduction in their transmission. Thus, this approach can describe transport from ballistic to diffusive (can be mapped to Boltzmann transport) and localized regimes [32-34]. Note that the details of the rough edges, strain fields, and relaxation that might develop could quantitatively affect $T_{ph}$, but we do not expect it to alter our qualitative conclusions about the nature of width-dependence. This is more adequately captured by the NEGF transport model we employ through the atomistic description of the edges.

As an example on how to identify the different transport regimes, in Fig. 1b we pick an energy $E = 0.13$ eV (~ in the middle of the spectrum) and plot the normalized transmission of those phonons versus channel width, $W$. We consider channels of lengths $L = 20$ nm (blue-squares), $L = 40$ nm (green-diamonds), and $L = 100$ nm (red-stars).

We then extract the exponent $\beta$ for the width-dependence of the normalized transmission as $W^\beta$. The shorter channel (blue line), has a rather weak width-dependence, especially for larger channel widths, approaching a $W^{0.6}$ behavior. For the narrower channels the exponent increases to $\beta = 4$. For the longer channel, (red line in Fig. 1b), the width-dependence becomes stronger especially for the narrow channels, where the exponent largely increases as $W^{21}$.

The different exponents reflect to different transport regimes, which can be identified by plotting the product of the normalized phonon transmission multiplied by the channel length, $T_{ph} \times L/W$ versus the channel length $L$, as shown in Fig. 1c for the phonons with energy $E = 0.13$ eV. Channels of two different widths $W = 1$ nm (solid line) and $W = 5$ nm (dashed line) are shown. Intuitively, for ballistic transport, the $T_{ph} \times L/W$ product increases linearly with channel length. For diffusive transport it remains constant. In the case of sub-diffusive transport it is reduced with $L$ [35, 36], and for localized transport it drops exponentially. Thus, from Fig. 1c, it can be deducted that the wider GNR channels (dashed line) are semi-ballistic for short channel lengths, reach the diffusive regime at lengths around $40$ nm $- 100$ nm, and the sub-diffusive regime for channel lengths beyond $\sim 100$ nm. Phonons in the ultra-narrow $W = 1$ nm GNRs (solid line) reach the localization regime for channels somewhat larger than $L \sim 10$ nm, in which case the phonon transmission is diminished.

We can now map these regimes to the exponents of Fig. 1b, which indicate $\beta < 1$ for semi-ballistic phonons, $\beta = 1$ for diffusive phonons, and $\beta > 1$ for sub-diffusive or localized phonons (see letters identifying the channels in the figures). Intuitively, ballistic transport will provide $\beta = 0$, in which case the line-edge roughness does not influence the phonon conductivity, and the normalized transmission is roughly independent of the channel's width. On the other hand, large $\beta$ signals localization, because as the width increases, localization weakens, which translates into a relatively large increase in the normalized transmission $T_{ph}/W$.

Using the $\beta$ extraction procedure as in the example of Fig. 1b, we now investigate the width-dependence of the normalized transmission for all phonons in the spectrum. As we previously describe in Ref. [25] within a given channel phonons of different frequencies can be ballistic, diffusive, or even localized. Thus, it follows that different phonon frequencies will have a different $\beta$ dependence as well. In Fig. 2 we extract the parameter $\beta$ of the normalized transmission in channels of widths $W = 1$ nm (blue-solid lines) and $W = 5$ nm (red-dashed lines). Results for three channel lengths $L = 20$ nm, 100 nm, and 500 nm are shown in Fig. 2a-c, respectively. In most of the spectrum, and for all channel lengths, $\beta$ is larger for the narrower GNRs because of the stronger roughness influence. For the narrower channel (solid-blue lines), the lowest $\beta$ value is observed for low frequency modes, whereas for energies above $E = 0.025$ eV $\beta$ increases drastically, and increases even more as the channel length increases. For the wider channels with $W = 5$ nm (dashed-red lines), the lowest $\beta$ value is also found at low energies. At higher energies $\beta$ resides around $\beta = 1$ (indicated by the dashed-black horizontal line), with a slight increase as the length of the channel increases (compare red lines in Fig. 2a to Fig. 2c).

The overall thermal conductivity of a channel and any experimentally determined $\beta$ value will include weighted contributions from all phonon modes in the spectrum. Figure 3 shows the width-dependence of the overall normalized thermal conductance (which has the same trend as conductivity) for the $L = 20$ nm (blue-squares), $L = 500$ nm (black-circles), and $L = 1000$ nm (red-triangles) channels in logarithmic scale, which allows for plotting straight lines of slope $\beta$. We plot the trends for low temperature $T = 20$ K (Fig. 3a), room temperature $T = 300$ K (Fig. 3b), and higher temperature $T = 450$ K (Fig. 3c), and label $W^\beta$ dependences in various places. The exponents $\beta$ vary significantly from $\beta = -0.45$ to $\beta = 2.6$ (we discuss the meaning of the negative $\beta$ below), indicating that the nature of thermal transport, from ballistic to localized, is channel geometry dependent.

We start our discussion with the room temperature results in Fig. 3b. For the shortest $L = 20$ nm channel (blue line), $\beta = 0.8$ is extracted independent of the channel's width, indicating that the overall phonon transport in this channel is quasi-ballistic towards diffusive, as one would have expected from such a short channel. As the channel length increases to $L = 500$ nm (black line) $\beta$ increases to $\beta = 1.45$ (for wider channels) and somewhat more to $\beta = 1.65$ as the width decreases down to $W = 1$ nm. These channels are clearly in the diffusive to sub-diffusive regime for all widths. For even longer channels, $L = 1000$ nm, although $\beta$ remains close to one for the wider channels ($\beta = 1.05$) indicating diffusive transport, as the width scales down to $W = 1$ nm, transport deviates from the diffusive to the sub-diffusive even the localized regime, and $\beta$ increases to $\beta = 2.6$. This behavior is in agreement with Fig. 2, where as the length of the channel increases (and the width decreases) more parts of the spectrum fall into localization (larger $\beta$'s), which increases the overall averaged $\beta$. The different transport regimes are also very well demonstrated by plotting the normalized phonon transmission multiplied by the channel length, $T_{ph} \times L/W$, versus $L$, as previously in Fig. 1c, but now $T_{ph}$ includes the contribution of all excited phonon modes. This is plotted in Fig. 4a for channel widths $W = 5$ nm (dashed line) and $W = 1$ nm (solid line). The increasing trend for short channels justifies the quasi-ballistic $\beta < 1$ in Fig. 3b, whereas the saturation with channel length for the wider GNR (dashed-red line) justifies the diffusive to sub-diffusive $\beta = 1 - 1.4$. The decreasing trend with length for the narrow GNR (solid-blue line) justifies the sub-diffusive/localized $\beta = 2.6$.

The $\beta$ exponents behavior we observe at room temperature in Fig. 3b, essentially remains unchanged for higher temperatures, at least up to $T = 450$ K, as indicated in Fig. 3c. A slight $\beta$ increase appears due to the fact that higher temperatures excite more high energy modes, which have a higher $\beta$ as we observed in Fig. 2. The change is small because room temperature already involves a large part of the spectrum anyway.

The situation changes, however, when we consider low temperatures, $T = 20$ K, as shown in Fig. 3a. Here only the low energy phonon modes with a very low $\beta$ are excited, and thus the overall $\beta$ is also significantly reduced. Interestingly, although long and narrow channels still show sub-diffusive behavior with $\beta = 1.9$ (red line – left side of Fig. 3a), shorter and/or wider channels have negative $\beta$ exponents. This means that the thermal conductivity can actually increase with decreasing channel width. This non-trivial effect is related to the increasing importance of the low frequency acoustic modes in determining the overall thermal conductivity. It was described in detail using simple analytical models in [19], but it is interesting that it is also verified by the more sophisticated NEGF method. The number of the acoustic modes does not scale with width as higher energy modes do and they are affected the least by line-edge roughness. Thus, as the width $W$ decreases, the conductivity $\kappa \sim K/W$ increases. Indeed, Fig. 4b shows a zoom in the low energy region of Fig. 2a in linear scale, indicating that for the narrow $W = 1$ nm channel most $\beta$ values are actually negative. Therefore, when the low energy phonon modes dominate transport, the overall exponent $\beta$ tends to decrease and even turn negative. This happens as the temperature decreases, but also interestingly as the length of the channel increases. In long channels, the high frequency modes fall deep into localization and do not contribute at all, which again increases the relative importance of the low-frequency modes. Thus, there is a competition between roughness reducing the conductivity (increasing $\beta$), but through this it increases the importance of the low-frequency phonons, which decreases $\beta$ again. This explains why in Fig. 3a and Fig. 3b, for the wider channels (right side), as the channel length increases one obtains a negative (or reduced) $\beta$ (red line).

An interesting observation here is that the value of $\beta$ in an experimental setup could point to the transport regime, in addition to extracting this information from the value of $\alpha$ in the length dependence $L^\alpha$ behavior. However, since the phonon transport regime, whether ballistic, diffusive, or localized, is determined by the channel length, then the width-dependence becomes a length-dependent property.

It is important at this point to examine how our diffusive results correlate with well-established macroscale models such as the Casimir model. The $\beta = 1$ value that corresponds to diffusive transport in our results, can be understood in simple terms by considering that the total phonon transmission, $T_{ph}$, can be expressed as:

$$T_{ph} = T_B \frac{\lambda}{\lambda + L} \qquad (1)$$

where $T_B$ is the ballistic transmission and $\lambda$ is the energy dependent phonon mean-free-path (mfp) for scattering. In the diffusive limit, $\lambda \ll L$, such that $T_{ph} = T_B \lambda / L$. The ballistic transmission $T_B$ is proportional to $W$, since the number of modes in the channel is affected almost linearly with $W$. Thus, in the diffusive limit, $T_{ph}/W \sim \lambda/L$. The dependence of the mfp on the GNR width, on the other hand, can be extracted by using the Casimir's model. This model considers that the phonon-boundary scattering rate is directly proportional to the phonon group velocity and inversely proportional to the channel's width as $1/\tau_B \sim v_g/W$. This makes the phonon-boundary mfp to be $\lambda = \tau_B \times v_g \sim W$. Thus, in simple terms, the overall dependence of the normalized phonon transmission, which determines the diffusive thermal conductivity trend, follows a linear dependence [31]. (Note that the quantities above can be extracted at individual phonon energies and branches, or be averaged/integrated over all energies and phonon branches).

In the channels we consider, with lengths over a few hundred nanometers and widths over $\sim 2$ nm (under diffusive transport conditions), we also observe the linear width-dependence, well aligned with the Casimir theory, as also shown by Carrete *et* al. for nanowires [37]. The Casimir theory for the fully diffusive case is predominantly employed for macroscale channels and roughness, but very frequently without much justification for low-dimensional channels as well. It is quite interesting to observe that although it is based on a very different logic, it provides the same width-dependence as the NEGF simulation results with atomistic roughness description even down to channels with width $W = 2$ nm. Note that this linear behavior is quite different from the boundary scattering dependence of electronic transport which follows $W^4$ for GNRs [29], and $W^6$ for nanowires or quantum wells [26-28].

Finally, we need to mention that anharmonic phonon-phonon interactions, which have been ignored in this work, tend to shift transport towards diffusion (especially at high temperatures) as they weaken the coherence that localization depends on [38, 39]. However, we believe that the influence of anharmonic interactions will have little qualitative and even less quantitative influence on our results. The reason is that the room temperature phonon-phonon interaction-limited mfp in graphene (and pristine GNRs) is considered to be $\lambda \sim 775$ nm [16, 40, 41], or sometimes even much longer up to $\lambda \sim 100$ μm [42-44], which is in the order of the channel lengths we consider, or longer. In fact, a recent theoretical work suggested that most if not all graphene experiments are carried out in the quasi-ballistic regime [45]. On the other hand, the roughness scattering-limited mfp is as low as $\lambda \sim 10$s of nanometers [25, 33], which makes phonon-boundary scattering the dominant scattering mechanism. At the higher temperature $T = 450$ K that we consider, qualitatively the $\beta$ values could be slightly shifted towards unity in the presence of anharmonic interactions.

In conclusion, using lattice dynamics and Non-Equilibrium Green's Function simulations, we have investigated the width-dependence of the thermal conductivity in ultra-narrow graphene nanoribbons in the presence of line-edge roughness. This is described in the form $W^{\beta}$, where $W$ is the width of the channel and $\beta$ corresponds to the roughness strength. We show that the exponent $\beta$ varies significantly depending on the nature of phonon transport, i.e. from quasi-ballistic ($\beta < 1$), to diffusive ($\beta = 1$), and to sub-diffusive and localized regimes ($\beta \gg 1$). Since the channel length determines the transport regime, however, we show that the width-dependence exponent $\beta$ becomes a channel length dependent property as well. Our results add to the current efforts in understanding heat flow at the nanoscale and could be relevant to low-dimensional systems in general, even beyond graphene nanoribbons.

**Acknowledgment:**
This work has received funding from the European Research Council (ERC) under the European Union's Horizon 2020 research and innovation programme (grant agreement No 678763)

# References

[1] N. Mingo, D. A. Broido, *Nano Lett.*, 5, 1221-1225, 2005.

[2] D. L. Nika, A. S. Askerov, and A. A. Balandin, *Nano Lett.*, 12, 3238-3244, 2012.

[3] C. W. Chang, D. Okawa, H. Garcia, A. Majumdar, and A. Zettl, *Phys. Rev. Lett.*, 101, 075903, 2008.

[4] L. Lindsay, D. A. Broido, and N. Mingo, *Phys. Rev. B*, 82, 115427, 2010.

[5] A. Balandin, *Nat. Materials*, 10, 569, 2011.

[6] X. Xu, L.F.C. Pereira, Y. Wang, J. Wu, K. Zhang, X. Zhao, S. Bae, C.T. Bui, R. Xie, J.T.L. Thong et al., Nat. Comm., 5, 3689, 2014.

[7] A. V. Savin, Y. S. Kivshar, and B. Hu, *Phys. Rev. B*, 82, 195422, 2010.

[8] Y. Wang, B. Qiu, and X. Ruan, *Appl. Phys. Lett.*, 101, 013101, 2012.

[9] Z. Aksamija and I. Knezevic, *Phys. Rev. B.*, 90, 035419, 2014.

[10] X. Ni, M. L. Leek, J.-S. Wang, and Y. P. Feng, B. Li, *Phys. Rev. B*, 83, 045408, 2011.

[11] J. Lan, J.-S. Wang, C. K. Gan, and S. K. Chin, *Phys. Rev. B*, 79, 115401, 2009.

[12] H. Karamitaheri, N. Neophytou, M. Pourfath, R. Faez, and H. Kosina, *J. Appl. Phys.*, 111, 054501, 2012.

[13] A. Cresti, N. Nemec, B. Biel, G. Niebler, F. Triozon, G. Cuniberti, and S. Roche, *Nano Research*, 1, 361-394, 2008.

[14] G. Iannaccone, Q. Zhang, S. Bruzzone, G. Fiori, *Solid-State Electronics*, 115, 213-218, 2015.

[15] M. Wang, N. Yang, and Z.-Y. Guo, *J. Appl. Phys.*, 110, 064310, 2011.

[16] D. Singh, J. Y. Murthy, and T. S. Fisher, *J. Appl. Phys.*, 110, 113510, 2011.

[17] S. Ghosh, W. Bao, D. L. Nika, S. Subrina, E. P. Pokatilov, C. N. Lau, and A. A. Balandin, *Nature Materials*, 9, 555-558, 2010.

[18] M. Bae, Z. Li, Z. Aksamija, P. N. Martin, F. Xiong, Z. Ong, I. Knezevic, and E. Pop, *Nat. Comm.* 4, 1734, 2013.

[19] H. Karamitaheri, N. Neophytou, and H. Kosina, *J. Appl. Phys.*, 115, 024302, 2014.

[20] M. Luisier, *J. Appl. Phys.*, 110, 074510, 2011.

[21] A. Hochbaum, R. Chen, R. Delgado, W. Liang, E. Garnett, M. Najarian, A. Majumdar, and P. Yang, *Nature*, 451, 163, 2008.

[22] S. Lepri, R. Livi, and Antonio Politi, *Phys. Rev. Lett.*, 78, 1896, 1997.

[23] B. Li and J. Wang, *Phys. Rev. Lett.*, 91, 044301, 2003.

[24] G. Wu and J. Dong, *Phys. Rev. B*, 71, 115410, 2005.

[25] H. Karamitaheri, M. Pourfath, H. Kosina, and N. Neophytou, *Phys. Rev. B*, 91, 165410, 2015.

[26] K. Uchida and S. Takagi, *Appl. Phys. Lett.*, vol. 82, no. 17, pp. 2916-2918, 2003.

[27] S. Jin, M. V. Fischetti, and T.-W. Tang, *J. Appl. Phys.*, 102, 083715, 2007.

[28] N. Neophytou and H. Kosina, *Phys. Rev. B*, 84, 085313, 2011.

[29] M. V. Fischetti and S. Narayanan, *J. Appl. Phys.*, 110, 083713, 2011.

[30] R. Saito, M. Dresselhaus, G. Dresselhaus, *'Physical Properties of Carbon Nanotubes'*, Imperial College Press, London, 1998.

[31] H. Karamitaheri, M. Pourfath, R. Faez, H. Kosina, *IEEE Transactions on Electron Devices*, 60, 2142, 2013.

[32] C. Jeong, S. Datta, and M. Lundstrom, *J. Appl. Phys.*, 111, 093708, 2012.

[33] H. Sevincli and G. Cuniberti, *Phys. Rev. B*, 81, 113401, 2010.

[34] S.G. Das and A. Dhar, *The European Physical Journal B*, 85, 372, 2012.

[35] B. Vermeersch, A.M.S. Mohammed, G. Pernot, Y.-R. Koh, A. Shakouri, *Phys. Rev. B*, 91, 085203, 2015.

[36] N. A. Gallo, M .I. Molina, *J. Phys. A: Math. Theor.*, 48, 045302, 2015.

[37] J. Carrete, L. J. Gallego, L. M. Varela, and N. Mingo, *Phys. Rev. B*, 84, 075403, 2011.

[38] S. Soleimani, S.B. Touski, M. Pourfath, *Appl. Phys. Lett.*, 105, 103502, 2014.

[39] R. Golizadeh-Mojarad, S. Datta, *Phys. Rev. B*, 75, pp. 081301, 2007.

[40] S. Ghosh, I. Calizo, D. Teweldebrhan, E. P. Pokatilov, D. L. Nika, A. A. Balandin, W. Bao, F. Miao, C. N. Lau, *Appl. Phys. Lett.*, 92, 151911, 2008.

[41] E. Munoz, J. Lu, and B. I. Yakobson, *Nano Lett.*, vol. 10, pp. 1652–1656, 2010.

[42] S. Mei, L. N. Maurer, Z. Aksamija and I. Knezevic, *J. Appl. Phys.*, 116, 164307, 2014.

[43] G. Fugallo, A. Cepellotti, L. Paulatto, M. Lazzeri, N. Marzari, and F. Mauri, *Nano Lett.*, 14, 6109, 2014.

[44] G. Barbarino, C. Melis, and L. Colombo, *Phys. Rev. B*, 91, 035416, 2015.

[45] C. Melis and L. Colombo, *Phys. Rev. Lett.*, 112, 065901, 2014.

Figure 1:

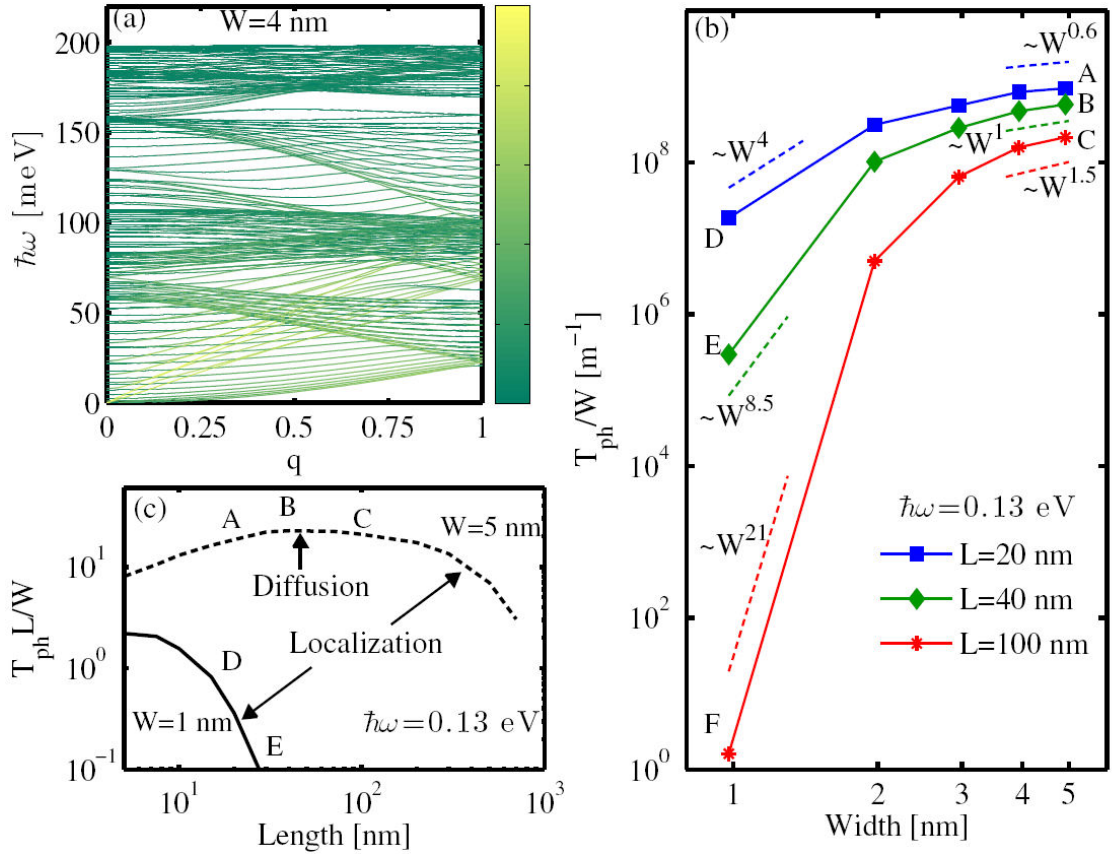

Figure 1 caption:

(a) Phonon dispersion for a $W$ = 5 nm GNR. The colormap shows the contribution of each phonon state to the total ballistic thermal conductance (light: largest contribution, dark: smallest contribution). (b) The width-normalized phonon transmission of rough nanoribbons of lengths $L$ = 20 nm (blue), $L$ = 40 nm (green), and $L$ = 100 nm (red) for phonon energy $E$ = 0.13 eV versus channel width, $W$. (c) The width-normalized phonon transmission times the channel length $T_{ph} \times L/W$ versus the channel length $L$ for phonons with energy $E$ = 0.13 eV. In (c) results for GNRs of width $W$ = 5 nm (dashed line) and $W$ = 1 nm (solid line) are shown. The different channels are labeled in 1b and 1c with the letters A-F.

Figure 2:

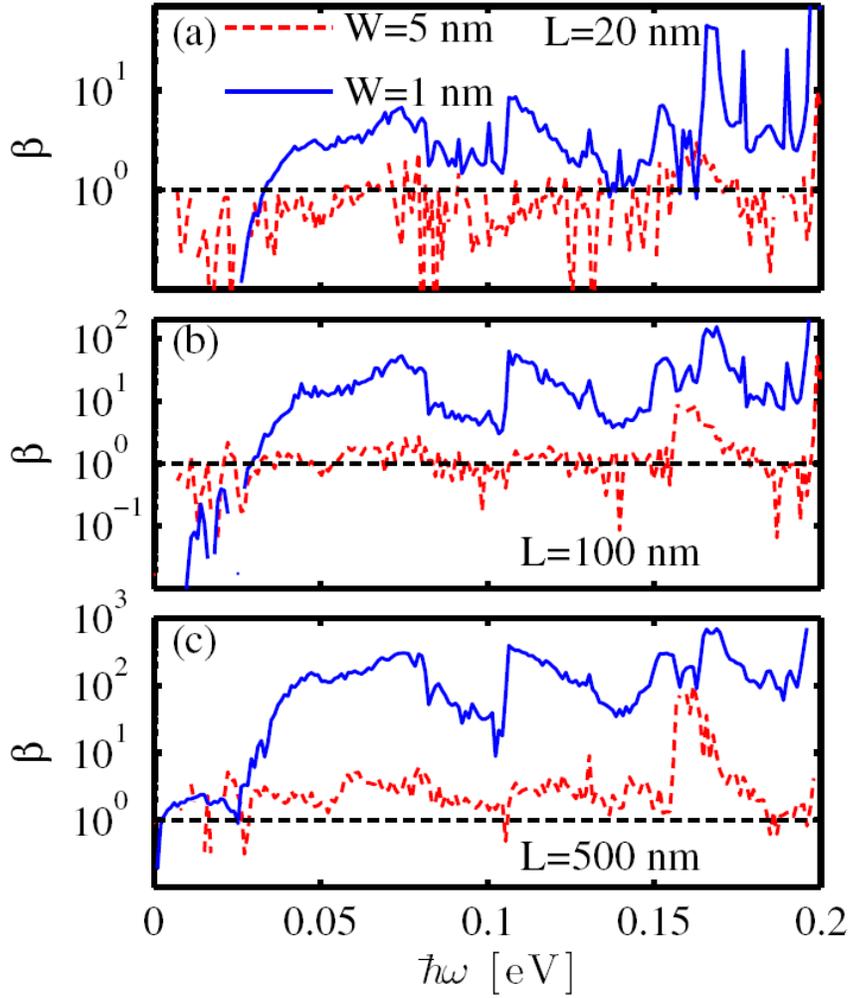

Figure 2 caption:

The width-dependent exponent $\beta$ for the width-normalized phonon transmission in the entire spectrum for rough GNRs of lengths $L = 20$ nm (a), $L = 100$ nm (b) and $L = 500$ nm (c). Results for GNRs of width $W = 5$ nm (dashed-red lines) and $W = 1$ nm (solid-blue lines) are shown. The dashed-black lines indicate the $\beta = 1$ value, which corresponds to diffusive transport.

Figure 3:

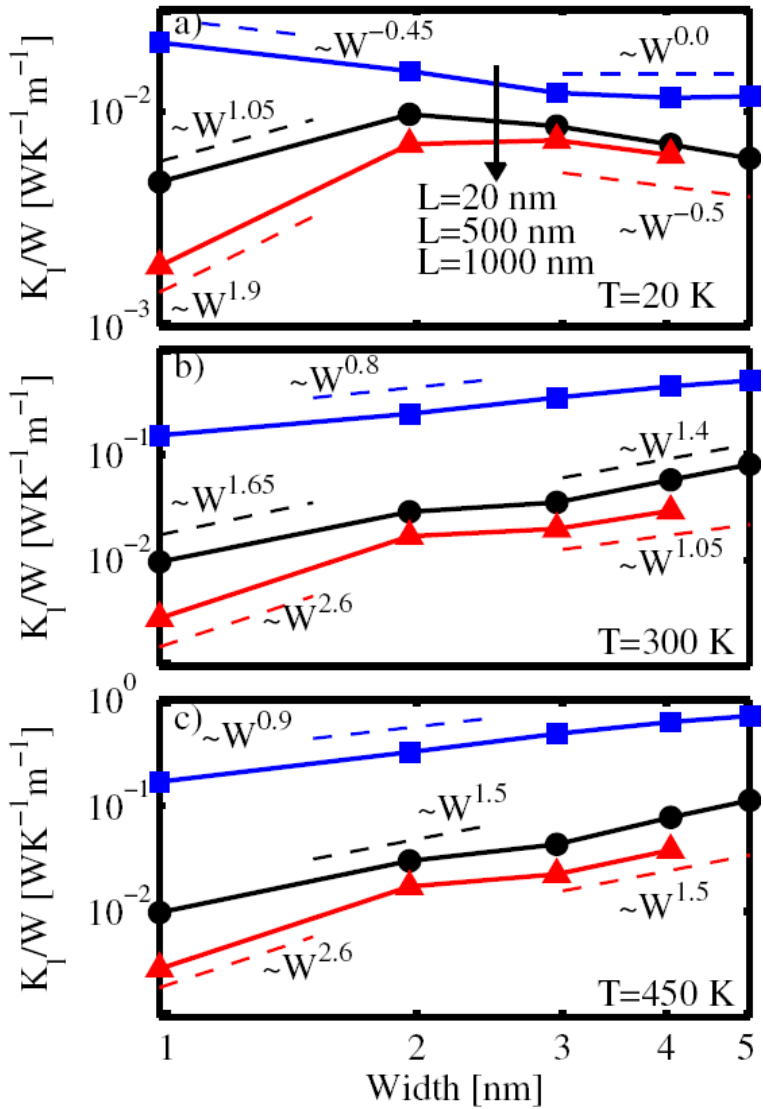

Figure 3 caption:

The width-normalized thermal conductance versus channel width $W$ for GNRs with lengths $L$ = 20 nm (blue-squares), $L$ = 500 nm (black-circles), and $L$ = 1000 nm (red-triangles). Results for temperatures (a) $T$ = 20 K, (b) $T$ = 300 K, and (c) $T$ = 450 K are shown. Characteristic exponents in the form $W^\beta$ are indicated for various parts of the straight lines that are fitted through the data, which vary from $\beta$ = -0.45 up to $\beta$ = 2.6.

Figure 4:

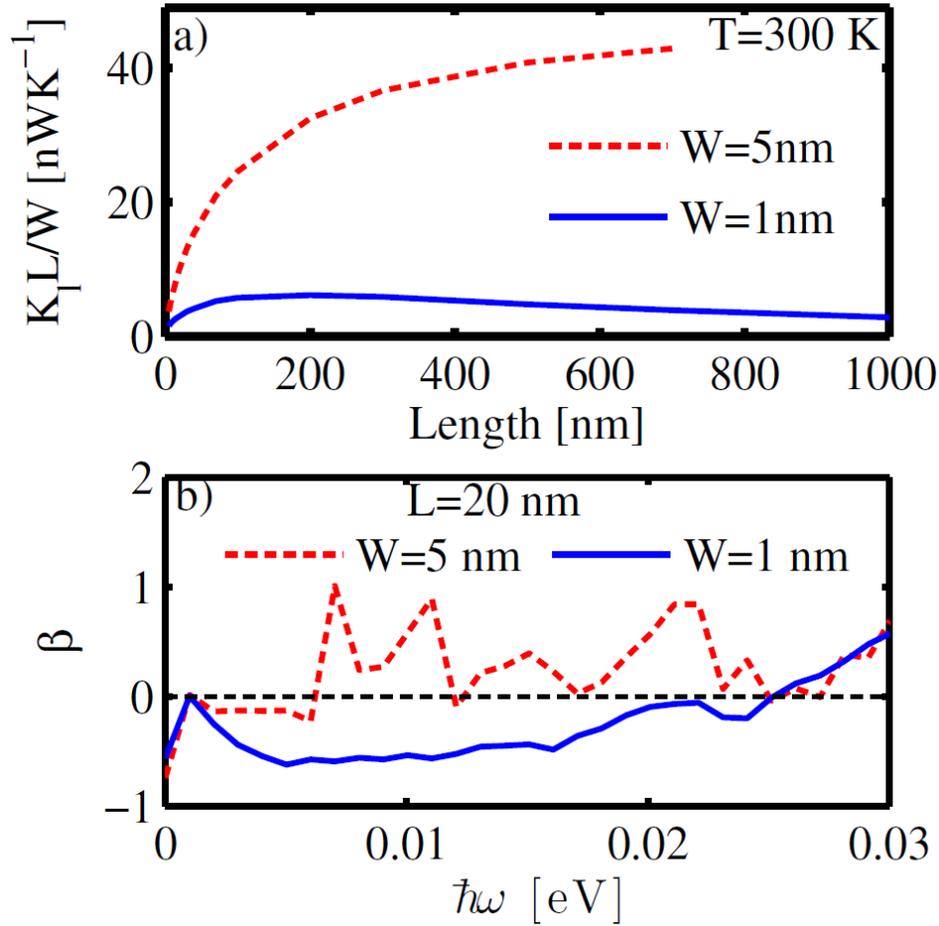

Figure 4 caption:

(a) The width-normalized thermal conductance times the channel length $K_l \times L/W$ versus the channel length $L$ at room temperature. Results for GNRs of width $W = 5$ nm (dashed-red lines) and $W=1$nm (solid-blue lines) are shown. (b) A zoom of the width-dependent exponent $\beta$ for the width-normalized phonon transmission in the low energy spectrum region of the rough GNR of length $L = 20$ nm (zoom of Fig. 2a). The dashed-black lines indicate the $\beta = 0$ value, which corresponds to ballistic transport.